\documentclass[prl, aps, superbib, tightenlines, twocolumn, floatfix]{revtex4}
\usepackage{amsmath, graphicx, dcolumn, bm}

\renewcommand{\vec}{\bf}
\renewcommand{\d}{{\rm d}}
\renewcommand{\i}{{\rm i}}
\begin{document}

\title{The decay of excited He from Stochastic Density-Functional
Theory: a quantum measurement theory interpretation}
\author{Neil Bushong}
\affiliation{
Department of Physics, University of California, San Diego, La Jolla,
CA 92093-0319}

\author{Massimiliano {Di Ventra}}
\email{diventra@physics.ucsd.edu}
\affiliation{
Department of Physics, University of California, San Diego, La Jolla,
CA 92093-0319}

\date{\today}

\begin{abstract}

Recently, time-dependent current-density functional theory has been
extended to include the dynamical interaction of quantum systems
with external environments [Phys. Rev. Lett. {\bf 98}, 226403
(2007)]. Here we show that such a theory allows us to study a
fundamentally important class of phenomena previously inaccessible
by standard density-functional methods: the decay of excited
systems. As an example we study the decay of an ensemble of excited
He atoms, and discuss these results in the context of quantum
measurement theory.
\end{abstract}

\maketitle

Density-functional theory (DFT), in both its ground-state and
time-dependent versions~\cite{hohenberg64, kohn65, runge84, ghosh88,
vignale96} has become the method of choice to study several
equilibrium and non-equilibrium properties of interacting
many-particle systems evolving under Hamiltonian dynamics.

There is, however, a large class of physical problems where the
dynamical interaction of a quantum system with an external
environment needs to be taken into account. To this class of {\em
open quantum systems} belongs also one of the most basic tenets of
Quantum Theory, namely the non-unitary evolution of a quantum state
due to the measurement by an apparatus. Non-unitary quantum
evolution also pertains to processes where the energy of the quantum
system relaxes into the degrees of freedom of a bath or reservoir,
like, e.g., the decay of exited systems. An understanding of such
processes from a microscopic point of view would represent a
substantial advancement in the study of open quantum systems.

To address the above issues, Di Ventra and D'Agosta (DD) have
recently proved~\cite{dagosta07} that given an initial quantum
state, and an operator $\hat V$ that describes the interaction of a
many-body system with an external environment, two external vector
potentials ${\bf A}({\bf r},t)$ and ${\bf A}'({\bf r},t)$ that
produce the same ensemble-averaged current density, $\overline{ {\bf
j}({\bf r},t)}$, must necessarily coincide, up to a gauge
transformation. The DD theorem thus extends the previous theorems of
dynamical DFT (that are one of its corollaries if $\hat V=0$), and
allows for the first-principles description of the dynamics of open
quantum systems via effective single-particle equations. This theory
has been named Stochastic Time-Dependent Current-DFT (Stochastic
TD-CDFT).

Here we apply the above theory to a previously inaccessible problem
via standard DFT methods: the decay of an ensemble of excited He
atoms. In addition, we interpret the results in the context of
quantum measurement theory by showing that the interaction with the
environment can be viewed as a continuous ``measurement'' of the
state of the system, thus making a connection between
density-functional theory and quantum measurement theory.

We consider two cases: 1) an ensemble of excited He$^+$ atoms, whose
dynamics can be directly compared with the one obtained from a
density-matrix approach. 2) An ensemble of neutral excited He atoms.
Our results reveal unexpected features of this problem, like the
dampening and modification of high-frequency oscillations during
energy relaxation of the ensemble towards its ground state.

The starting point of Stochastic TD-CDFT is the stochastic equation
of motion of an auxiliary Kohn-Sham (KS) Slater determinant
$\Psi^{KS}$ built out of single-particle KS states $\phi_{\alpha}$
(atomic units are used throughout this paper)
\begin{equation}\label{evolutionKS}
\begin{split}
\partial_t \Psi^{KS}(t) = &-\i\sum_i\hat H_i^{KS}(t) \Psi^{KS}(t)
    - \frac{\tau}{2} \hat V^\dagger \hat V \Psi^{KS}(t) \\
& + \ell(t)\hat V\Psi^{KS}(t) \, ,
\end{split}
\end{equation}
where
\begin{equation}
\hat H_i^{KS}(t) = \frac{
    \left[
        \hat p_i + {\bf A}({\hat r_i},t)/c
        + {\bf A}_{xc}(\hat r_i,t)/c
    \right ]^2}{2}
+ \hat V_H(\hat r_i,t) \, ,
\label{hKS}
\end{equation}
with ${\bf A}({\hat r_i},t)$, an arbitrary external vector potential,
${\bf A}_{xc}[\overline{ {\bf j}({\bf r},t)}, |\Psi_0\rangle,
\hat{V}]$ the exchange-correlation vector potential (which is a
functional of the average current $\overline{\bf j}$, the initial
condition $|\Psi_0\rangle$, and the operator $\hat V$), and $\hat
V_H({\bf r},t)$ the Hartree potential. The quantity $\tau$ has
dimensions of time. Without loss of generality the stochastic process,
$\ell(t)$, is chosen such that it has both zero ensemble average and
$\delta-$autocorrelation, i.e.~\footnote{In the calculations of this
paper, we use small but finite time steps of duration $\Delta t$. The
mean and autocorrelation of $\ell(t)$ are thus given by
$\overline{\ell(t_i)} = 0$ and $\overline{\ell(t_i) \ell(t_j)} =
\delta_{i,j} \tau/\Delta t$, respectively, where $t_i$ and $t_j$ are
arbitrary times, and $\delta_{i,j}$ is the Kronecker delta.  We have
chosen the probability distribution $P(\ell)$ of $\ell(t)$, to be
Gaussian, so that $P(\ell) = \sqrt{\Delta t / 2\pi\tau} \,{\rm
exp}(-\ell^2 \Delta t / 2 \tau)$.},
\begin{equation}
\overline{\ell(t)}=0; \;
\overline{\ell(t)\ell(t')}=\tau\delta(t-t') \, ,
\label{eq:autocorr}
\end{equation}
where the symbol $\overline{\cdots}$ indicates the average over a
statistical ensemble of identical systems all prepared in the same
initial quantum state $|\Psi_0\rangle$. For the particular choice of
bath operator we will make in this paper (Eq.~(\ref{eq:defv})),
which acts on single-particle states only, the stochastic
equation~(\ref{evolutionKS}) is simply
\begin{equation}
\partial_t \phi_\alpha(t) = - \i \hat H_{\rm KS} \phi_\alpha(t)
    -\frac{\tau}{2} \hat V^\dagger \hat V \phi_\alpha(t)
    + \ell(t) \hat V \phi_\alpha(t) \, ,
\label{eq:tdsle}
\end{equation}
where $\alpha$ contains also the spin degrees of freedom.

The use of a stochastic Schr{\"o}dinger equation in the context of
DFT, and not of an equation of motion for the density matrix, is
because in DFT the KS Hamiltonians depend on the density (and/or the
current density), and therefore they are, in general, different for
the different elements of the ensemble. This does not generally
guarantee a closed equation of motion for the single-particle KS
density matrix of the mixed state~\cite{dagosta07}.

As mentioned previously, our aim is to describe the decay of excited
electrons bound to a He nucleus.  The electrons are prepared in some
initial excited state, and evolve into the ground state as a result of
the stochastic interaction with an environment, that, quite generally,
can be thought of as a boson field. The precise form of the operator
$\hat{V}$ which causes this behavior would, in general, depend on the
detailed model of the environment. Here we choose the simplest
possible operator, whose matrix elements are~\footnote{In our
numerical work we have used a position basis so that the action of
this operator in this basis is $\langle {\vec r} | V \phi(t)
\rangle = \sum_i \langle {\vec r} | \epsilon_i \rangle \sum_j
\langle \epsilon_i | V | \epsilon_j \rangle \langle \epsilon_j |
\phi(t) \rangle = \langle {\vec r} | E_0 \rangle \frac{1}{\sqrt{\tau
t_{\rm d}}} \sum_{j=1}^{j<M} \langle \epsilon_j | \phi(t) \rangle$.}
\begin{equation}
\langle \epsilon_i | \hat V | \epsilon_j \rangle =
\left\{
    \begin{array}{ll}
            1/\sqrt{\tau t_{\rm d}}
            & \;\;\;\textrm{if $i = 0$, and $0 < j < M$} \\
            0 &  \;\;\;\textrm{otherwise,} \\
    \end{array}
\right.
\label{eq:defv}
\end{equation}
where $| \epsilon_i \rangle$ is an eigenstate of the ground-state KS
Hamiltonian in the absence of the interaction with the bath,
$\epsilon_i$ is the corresponding eigenvalue, and the upper limit $M$
is a given integer representing the number of states we keep in the
simulation.  (In the present case we have kept $M=15$ states.) The
parameter $t_{\rm d}$ gives the timescale over which the decay will
occur, with larger values of $t_{\rm d}$ leading to longer decay
times~\footnote{In Ref. \onlinecite{vankampenbook}, a similar operator
for a two-state system is provided to describe energy relaxation and
decoherence.}. In the following we have chosen $t_d=1$ fs. The
operator $\hat V$ defined this way ensures that the stochastic
Schr{\"o}dinger equation~(\ref{eq:tdsle}) is independent of the
magnitude of $\tau$.

Clearly, the above operator reduces the projection of a
wave-function from the states $\{|\epsilon_1\rangle,
|\epsilon_2\rangle \ldots |\epsilon_{M-1}\rangle \}$, and increases
the projection onto the ground state $|\epsilon_0\rangle$.
Physically, it describes energy relaxation and dephasing.

\begin{figure}
\includegraphics[width=3.2in]{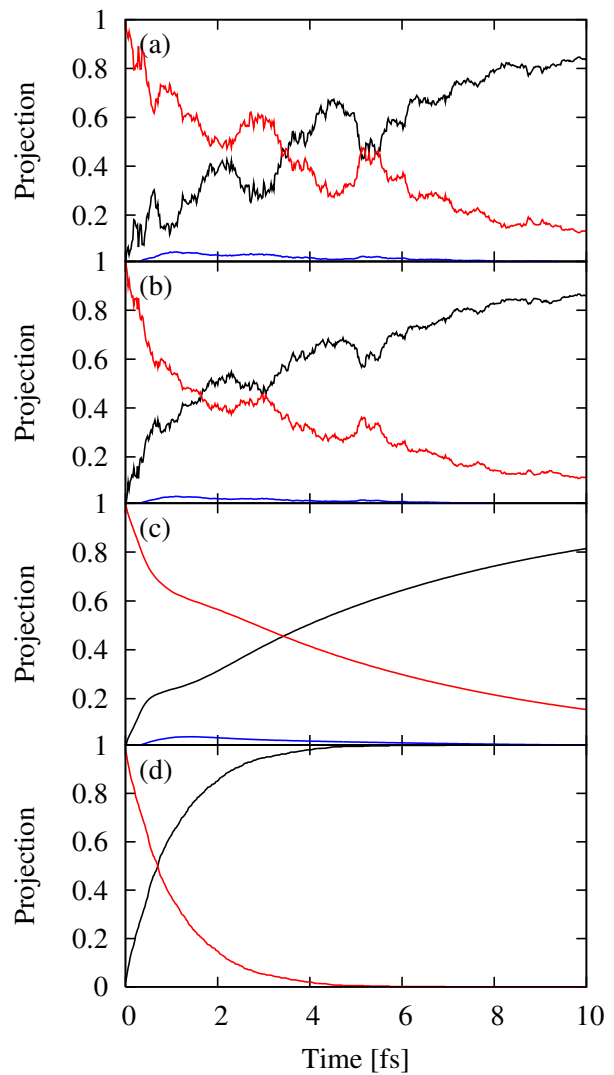}
\caption{(Color online) \emph{Panel~(a):} Stochastic evolution of the
projections $P_i$ onto the unperturbed states $\epsilon_i$, for
He$^+$, as a function of time. The black, red, and blue curves
correspond to $P_0$, $P_1$, and $P_2$, respectively.  Note that the
projections $P_3$ and $P_4$ coincide with $P_2$, since all three
states have $p$ symmetry. All other projections are vanishingly small.
\emph{Panel~(b)} Same as for panel~(a), but averaged over five
different runs, each with a different seed for the random number
generator. \emph{Panel~(c):} Same as for panels~(a) and (b), except
that the dynamics were calculated using the Lindblad master equation
(\ref{eq:sl_lindblad}). \emph{Panel~(d):} Same as for panels~(a)-(c),
except the dynamics were calculated using the wave-packet collapse
methodology of equation (\ref{eq:mcwf}).}
\label{fig:he_plus_decay}
\end{figure}

The stochastic Schr{\"o}dinger equation~(\ref{eq:tdsle}) preserves the
ensemble-averaged wave-function normalization~\cite{dagosta07}.
However, the normalization is not necessarily satisfied for any
particular realization of $\ell(t)$. In order to reduce the number of
dynamical calculations to perform the ensemble average, we have
explicitly re-normalized $|\phi \rangle$ at every time step. As we
will see below, with this approximation the decay into the ground
state is evident even after a single realization of
$\ell$~\footnote{Nevertheless, this creates slight differences in the
long-time limit of the dynamics with respect to the exact solution
(see, e.g, Fig.~\ref{fig:he_plus_decay}).}.

We begin by considering the behavior of an ensemble of He$^+$ ions
interacting with the environment represented by the
operator~(\ref{eq:defv}). We prepare the system with all ions in the
ensemble in the 2s state, denoted by $|\epsilon_1 \rangle$, and then
let the electrons evolve in time according to equation
(\ref{eq:tdsle})~\footnote{We represent the He nucleus with a simple
$2/r$ potential.  We integrate out the singularity at the origin using
a method similar to the Ewald method~\cite{alavi00, bylaska02}.  We
use the Hockney method to calculate the potential of an isolated
system~\cite{hockney70}.  The supercell is a cube of length 16.93 \AA,
and the grid spacing is 0.239 \AA.  We use the Chebyshev method of
constructing the propagator~\cite{tal-ezer84,leforestier90}, and we
use a time step of 0.02 fs.}.  Panel~(a) of
Fig.~\ref{fig:he_plus_decay} gives the projections $P_i(t) = | \langle
\epsilon_i | \phi(t) \rangle |^2$, as a function of time for one
particular realization of $\ell(t)$.  We see that the projection
$P_0(t)$ onto the ground state approaches one as time evolves, while
the projections onto other states approach zero, indicating energy
relaxation into the ground state.  In order to demonstrate that this
behavior is not due to the particular choice of seed in our random
number generator, we also plot the projections $P_i(t)$ averaged over
5 different simulations with different seeds. One can clearly see that
the fluctuations in panel~(b) of Fig.~\ref{fig:he_plus_decay} are
reduced in comparison to the fluctuations in panel~(a).

For the single-electron case of He$^+$, we can analytically treat
the ensemble average over all realizations of $\ell(t)$ by
considering the density matrix of this mixed state. Using equations
(\ref{eq:tdsle}) and (\ref{eq:autocorr}), it can be
shown~\cite{vankampenbook,dagosta07} that the resultant density
matrix ${\hat \rho}$ evolves according to the Lindblad master
equation
\begin{equation}
\frac{\d {\hat \rho}}{\d t} = -\i [H_{\rm KS}, {\hat \rho}]
    + \tau V {\hat \rho} V^\dagger
    - \frac{\tau}{2} {\hat \rho} V^\dagger V
    - \frac{\tau}{2} V^\dagger V {\hat \rho} \, .
\label{eq:sl_lindblad}
\end{equation}
In panel~(c) of Fig.~\ref{fig:he_plus_decay} we plot the matrix
elements $\langle \epsilon_i| {\hat \rho} | \epsilon_i \rangle$
showing the same behavior obtained with
Eq.~(\ref{eq:tdsle})~\footnote{Note that the density matrix procedure
we have used here does not take into account the additional
re-normalization that we apply in order to speed convergence in the
calculations with Eq.~(\ref{eq:tdsle}). The close correspondence
between panels~(b) and (c) in Fig.~\ref{fig:he_plus_decay}
demonstrates that our forced normalization procedure does not cause
significant deviation from the true average behavior of the system.}.

We now discuss this result in terms of measurement theory. It is
well-known that is possible to interpret the interaction with an
environment as a continuous ``measurement'' of the state of the system
-- or, equivalently, of the state of the environment -- with
consequent non-unitary wave-packet reduction~\cite{zurek03,
schlosshauer04}.  We can make this point even clearer by assuming that
every time the system interacts with the environment it emits a boson
excitation (whether a photon or a phonon) and thus there is a finite
probability $\d p = \d t | \langle \epsilon_1 | \phi \rangle |^2/t_d$
that the the emitted excitation be detected by an
apparatus~\cite{dalibard92}. Upon detection of this excitation, the
wave-function $|\phi \rangle$ {\em collapses} to the ground
state $|\epsilon_0 \rangle$. This is the well-known postulate of
wave-packet reduction.

We can write the above in the form of a Schr{\"o}dinger-type
equation of motion that includes a stochastic variable $\gamma (t)$,
which has a probability distribution uniformly distributed between 0
and 1.  If $\gamma > \d p$, an emitted excitation is not detected,
while if $\gamma < \d p$, the emitted excitation is detected, and
the wavefunction collapses to the ground state.  That is, during a
small time $\Delta t$, $|\phi \rangle$ evolves according to
\begin{equation}
\begin{split}
|\phi(t + \Delta t) \rangle
    & = {\rm e} ^{- \i H_{\rm KS} \Delta t} \theta(\gamma(t) - \d p)
        |\phi(t) \rangle \\
    & \quad + \theta(\d p - \gamma(t)) |\epsilon_0 \rangle \, ,
\end{split}
\label{eq:mcwf}
\end{equation}
where $\theta(x)$ is the Heaviside step function~\footnote{It can be
shown~\cite{dalibard92} that by defining an operator $\hat S$ such
that $\hat S |\epsilon_1 \rangle = |\epsilon_0 \rangle$, the density
matrix $\hat \rho$ associated with Eq.~(\ref{eq:mcwf}) satisfies the
Lindblad equation~(\ref{eq:sl_lindblad}) with $\hat S$ replacing
$\hat V$.}. A similar approach has been used by Dalibard \emph{et
al.} in the context of quantum optics~\cite{dalibard92}.

In Fig.~\ref{fig:he_plus_decay}(d), we plot the results from the
time evolution of equation (\ref{eq:mcwf}) for the problem of He$^+$
relaxation, where we have considered only the ground state and the
first excited state. Starting with $|\phi(t=0) \rangle =
|\epsilon_1\rangle$, we evolved equation (\ref{eq:mcwf}) in time for
1000 different realizations of $\gamma(t)$, and found the average
value of $| \langle \epsilon_i | \phi(t) \rangle |^2$, which we have
denote by $P_i$.  Note that each individual wave-function starts in
the excited state, and then suddenly drops to the ground state the
first time that $\gamma(t) < \d p$.  This wave-packet reduction
occurs at a different time for each run, and the ``remaining''
excited states become exponentially less likely as time goes on;
therefore, the average curve approaches an exponential. By comparing
Fig.~\ref{fig:he_plus_decay}(d) with Fig~\ref{fig:he_plus_decay}(c)
we see that the non-unitary wave-packet reduction evolution is
qualitatively similar to that provided by the stochastic
Schr{\"o}dinger equation~(\ref{eq:tdsle}), the difference being in
the fact that we have included only two states in the analysis of
Eq.~(\ref{eq:mcwf}). This equivalence therefore illustrates a point
of contact between the stochastic Schr{\"o}dinger
equation~(\ref{eq:tdsle}) and quantum measurement theory: the
environment ``measures'' the state of the system, and, as a result,
the wave-function is modified in a non-unitary way.

We conclude by discussing the decay of neutral He, where a closed form
for the KS single-particle Lindblad equation cannot be obtained. We
prepare the system in such a way that both electrons (with spin
$\sigma$) are in the first excited state of the ground-state
Hamiltonian, $|\phi_\sigma (t=0) \rangle = |\epsilon_1\rangle$. This
means that the Pauli exclusion principle is automatically satisfied by
our environment operator~(\ref{eq:defv}). For this case we compare the
stochastic evolution with the one in which $\hat V=0$~\footnote{We use
the local density approximation to the scalar exchange-correlation
potential~\cite{kohn65, runge84}, as derived by Ceperley and
Alder~\cite{ceperley80} and parametrized by Perdew and
Zunger~\cite{perdew81}.  For the current-density functional, we use
the interpolation formula of Conti and Vignale~\cite{conti99}.}.

\begin{figure}
\includegraphics[width=3.2in]{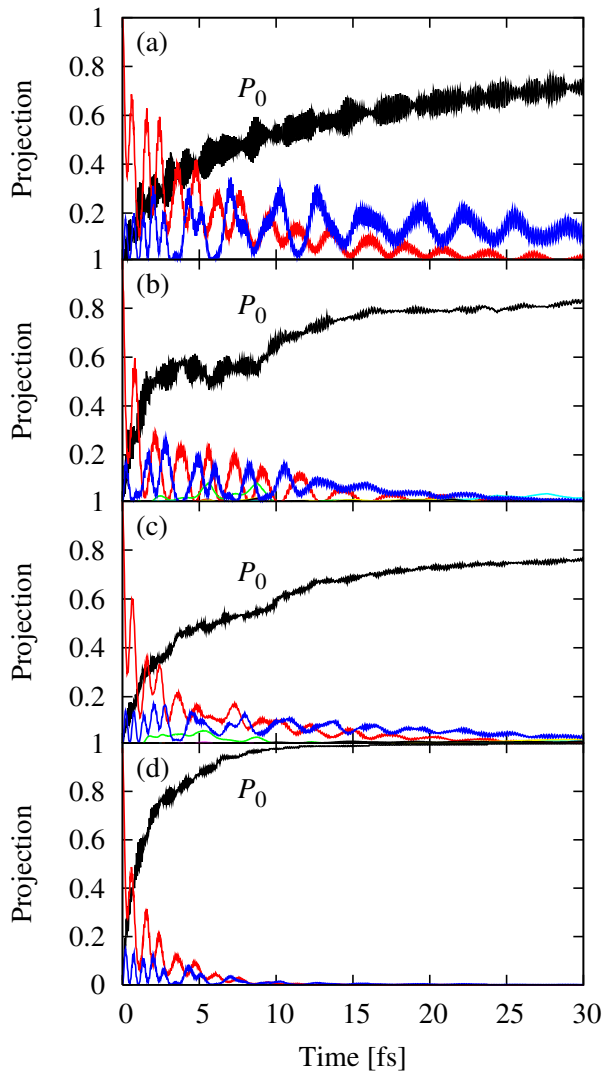}
\caption{(Color online) \emph{Panel~(a):} Projections $P_i = | \langle
\epsilon_i | \phi \rangle |^2$ for neutral He, as a function of time,
for the case where the stochastic terms are not included (unitary
evolution). The black, red, and blue curves correspond to $P_0$,
$P_1$, and $P_8$, respectively.  \emph{Panel~(b):} Same for as
panel~(a), but with the inclusion of the interaction with the
environment.  \emph{Panel~(c):} Same for as panel~(b), but averaged
over five different runs, each with a different seed for the random
number generator.  \emph{Panel~(d):} Same as for panels~(a)-(c),
except the dynamics were calculated using the wave-packet collapse
methodology of Eq.~(\ref{eq:mcwf}).}
\label{fig:he_decay}
\end{figure}

In Fig.~(\ref{fig:he_decay}), we plot the projections $P_i = |\langle
\epsilon_i | \phi \rangle|^2$ for the unitary evolution (panel~(a)),
as well as the projections for the non-unitary evolution for one
realization of $\ell(t)$ (panel~(b)) and averaged over five different
realizations of $\ell(t)$ (panel~(c)).  As expected, in the presence
of the environment, the projection onto the ground state $|\epsilon_0
\rangle$ approaches 1, while the occupations of other states are
suppressed as time goes on.  Here, however, we also note another
effect of the interaction. Fig.~\ref{fig:he_decay}(a) illustrates
that, in the unitary evolution the projections $P_i$ oscillate in
time. This oscillatory behavior reflects the motion of the electrons
as they alternately fall toward the nucleus, and then rebound
outward~\footnote{Sugino \emph{et al.} report similar high-frequency
oscillations while studying the dipole moment of an isolated aluminum
dimer~\cite{sugino99}.}.  Interaction with the environment has the
effect of not only dampening these oscillations but also of modifying
their frequency, the details of which vary depending on the particular
realization of $\ell(t)$.  The introduction of the bath mediates new
transitions for the single-particle wavefunction $|\phi \rangle$.

Similarly to the case of He$^+$, we can make a connection with quantum
measurement theory and study the decay of neutral He using
Eq.~(\ref{eq:mcwf}). This is illustrated in
Fig.~\ref{fig:he_decay}(d).  Here, however, we observe an important
qualitative difference. In the wave-packet reduction formalism
described by Eq.~(\ref{eq:mcwf}), after the apparatus ``detects'' the
excitation, the system immediately collapses onto the ground state
$|\epsilon_0 \rangle$, {\em irrespective} of the interactions of the
other states with the environment. In contrast, the non-unitary
evolution of the stochastic Schr{\"o}dinger Eq.~(\ref{eq:sl_lindblad})
involves a constant process of {\em self-consistent} interaction with
the reservoir. This implies that the frequency of small oscillations
is unchanged in the wave-packet reduction formalism, while they change
in time during dynamical interaction with the environment as described
by Eq.~(\ref{eq:sl_lindblad}). Clearly, one could introduce this
effect into the wave-packet reduction Eq.~(\ref{eq:mcwf}), but at a
non-trivial complexity cost, while the stochastic Schr{\"o}dinger
Eq.~(\ref{eq:sl_lindblad}) contains it naturally.

In summary, we have used Stochastic TD-CDFT to describe the
interaction of an excited quantum system (He) with an external
environment and its consequent decay into the ground state; a
problem previously inaccessible via standard DFT methods. We have
made a connection of this open quantum problem with quantum
measurement theory thus showing that Stochastic TD-CDFT may find
applications in quantum information theory of realistic systems.

We thank Y. Pershin for useful discussions and for pointing out
Ref.~\cite{dalibard92} to us. This work has been supported by the
Department of Energy grant DE-FG02-05ER46204.



\end{document}